\documentclass{emulateapj} 
\usepackage{apjfonts}

\usepackage{amsmath}
\usepackage{verbatim} 

\newcommand{\icarus}{{\it Icarus}}

\shortauthors{GOULD ET AL}

\shorttitle{EVIDENCE FOR CORE ACCRETION}

\begin{document}

\title{Kepler planet candidates consistent with core accretion}

\author{Andrew Gould, Jason Eastman}\affil{The Ohio State University, Columbus, OH 43210; gould@astronomy.ohio-state.edu}

\begin{abstract}

We show that the distribution of Kepler candidate planets from Borucki
et al.  is consistent with the predictions of the core accretion
model.

\end{abstract}

\maketitle

\begin{figure}[h]
  \begin{center}
    \includegraphics[width=3.25in]{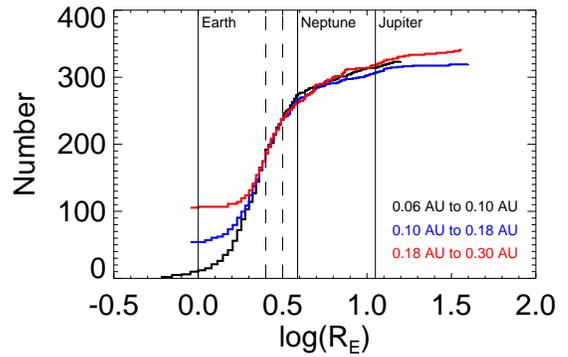} 
    \caption{The cumulative distribution functions of planetary
    radius, in Earth radii, of the Kepler planet candidates. The three
    lines are the distribution functions in different bins equally
    spaced in log semi-major axis between 0.06 AU and 0.3 AU. Each
    distribution was normalized to the first bin by matching the slope
    and intercept of the best fit line between $0.4 < \log{(R_E)} <
    0.5$ (boundaries are vertical dashed lines). The solid vertical
    lines, from left to right, show the radius of Earth, Neptune, and
    Jupiter.}
    \label{fig:cumulative}
  \end{center}
\end{figure}

In the distribution of Kepler candidate planets \citep{borucki11}, we
see a clear break in the number of planets larger than about a Neptune
radius, which is a generic prediction of core accretion theory
\citep{pollack96,ida04a}. In Figure \ref{fig:cumulative}, we show
cumulative distributions of the Kepler planet candidates in three
bins, equally spaced in log semi-major axis to show that the break is
present in each. We scale each distribution function to match the
slope and zero point of a best fit line between the two dashed lines
of the first bin in log semi-major axis to compare the distributions
more easily. We use the interval $0.4 < \log{(R_E)} < 0.5$ to scale,
where our lower bound corresponds to the smallest single-transit event
(KOI 364.01) and our upper bound is well before the observed break. We
see that all distributions show a very similar behavior with a clear
break at Neptune radius, as predicted by the core accretion model.

To rescale our distributions, the best fit slopes for the three bins
in semi-major axis, from smallest to largest are 530, 584, 369 stars
per $\log(R_E)$, and the intercepts are -26, -88, and -91 stars.

%\bibliography{references}{}
%\bibliographystyle{apj} 

%\begin{comment}

%\end{comment}

\end{document}